# HEP Analysis Facility – An Approach to Grid Computing

## Vivek Chalotra, AIE*, Anju Bhasin, Anik Gupta, Sanjeev Singh Sambyal

HEP Analysis Facility is a cluster designed and implemented in Scientific Linux Cern 5.5 to grant High Energy Physics researchers one place where they can go to undertake a particular task or to provide a parallel processing architecture in which CPU resources are shared across a network and all machines function as one large supercomputer.

## INTRODUCTION

Protons, electrons, neutrons, neutrinos and even quarks are often featured in news of scientific discoveries. All of these, and a whole "zoo" of others, are tiny sub-atomic particles too small to be seen even with the microscopes. While molecules and atoms are the basic elements of familiar substances that we can see and feel, we have to "look" within atoms in order to learn about the "elementary" subatomic particles and to understand the nature of our Universe. The science of this study is called Particle Physics, Elementary Particle Physics or sometimes High Energy Physics (HEP). High processing speed, large memory space and big storage are the main requirements of the HEP research applications like ROOT, ALIROOT, GEANT etc. HEP Analysis Facility is a cluster designed in Scientific Linux Cern 5.5.

*Department of Physics, Baba Saheb Ambedkar Road, University of Jammu, Jammu-180006, India.*

## OVERVIEW OF THE CLUSTER

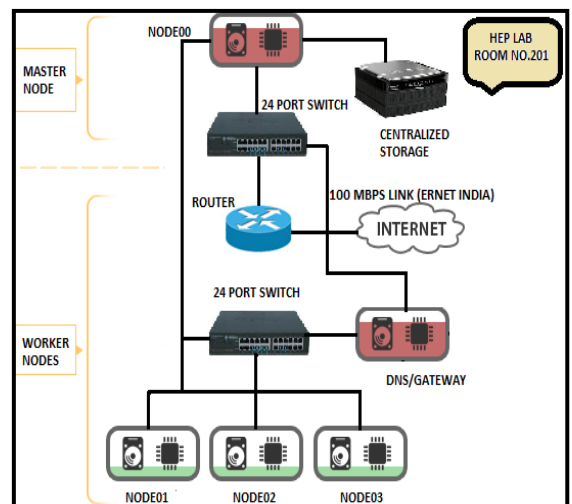

**Fig.1:** Overview of the Cluster

In this cluster our approach is to design and incorporate a middleware program, so that the HEP Analysis Facility cluster can work like a grid and the users on a TCP/IP network or the internet can run their parallel processing based research programs more effectively. This program gets information of available computing resources from all the worker nodes and automatically assigns the job to the node having most available computing resources like ACR (Available CPU Resources) and AMR (Available

Memory Resources). It also keeps track of each computers IP address and the time the job was submitted to the master.

**EQUIPMENTS USED**

| S.NO. | Hardware | Qty. |
|---|---|---|
| 1. | HP Proliant ML150 Server | **5Nos.** |
| 2. | Cisco 1841 series router | **1No.** |
| 3. | Cisco Catalyst 2950 switch | **1No.** |
| 4. | D-Link DES - 1008D switch | **1No.** |
| 5. | 5KVA UPS(2 hrs. backup) | **1No.** |
| 6. | Storage Device 8TB | **1No.** |

**Table 1:** Equipments Used

**CLUSTER COMPONENTS**

1. Master Node (node00).
2. Worker Nodes(node01, node02 & node03).
3. Storage Element (/Jugrid).
4. 100 Mbps internet lease line link provided by Ernet India (ISP).

**INSTALLATION OF NODES**

1. Scientific Linux 5.5 is the prerequisite for all the nodes in the cluster.
2. Partition table of worker nodes(Using LVM):

   /boot  =  500MB
   /home  =  100GB
   /      =  100GB
   swap   =  8GB

3. Partition table of master node(Using LVM ):

   /boot  =  500MB
   /home  =  100GB
   /      =  100GB
   swap   =  16GB

**NETWORK SETUP**

Setup TCP/IP networking by giving IP addresses to all the machines. For example:

Cisco Switch    :   144.x.x.2
Master Node     :   144.x.x.3(eth0),
                    192.x.x.1(eth1)
Worker Node 1   :   192.x.x.2(eth0)
Worker Node 2   :   192.x.x.3(eth0)
Worker Node 3   :   192.x.x.4(eth0)
Storage         :   192.x.x.5(eth0)
DNS             :   144.x.x.4(eth0)
                    192.x.x.6(eth1)
D-Link Switch   :   192.x.x.7

Configure DNS for forward & reverse mapping of all the nodes and make it an internet gateway so that all the machines in the cluster can have internet connectivity. Then write a SSH login script to make entire cluster nodes password free among each other.

## CENTRALIZED STORAGE

Centralized Storage of 8TB is mounted as /Jugrid in the master node and exported to all the worker nodes by editing the /etc/exports file like the following:

*/jugrid    node01(rw,rsync)    node02(rw,rsync)    node03(rw, rsync).*

Below are the commands used to mount it from the worker nodes:

*# mount node00:/Jugrid /home*

*# echo "192.x.x.1:/Jugrid /home nfs defaults 0 0" >> /etc/fstab*

## CREATING USERS

Create some users in /Jugrid(Centralized Storage) using the below command and copy the /etc/passwd, /etc/group & /etc/shadow files of the master into the workers so that the home directories of all the users become completely transparent to them:

*useradd -gusers -Gusers -s/bin/shell -pxxxx -d/Jugrid/username -m username*

## INSTALLATION OF HEP RESEARCH APPLICATIONS

Before the installation of Root and Aliroot in the centralized storage (/Jugrid) we have to set environment variables in /etc/bashrc file of the master node and then install the HEP applications like the following:

### Root Installation:-

mkdir /Jugrid/alice

cd /Jugrid/alice

svn co https://root.cern.ch/svn/root/tags/v5-26-00b root_v5.26.00

ln -s root_v5.26.00 root

cd $ROOTSYS

./configure

make

make install

Installation complete.

Give **root** command to run the application.

### Aliroot Installation:-

svn co https://alisoft.cern.ch/AliRoot/branches/v4-18-Release AliRoot_v4.18-Release

ln -s AliRoot_v4.18-Release AliRoot

cd $ALICE_ROOT

make

Installation complete.

Give **aliroot** command to run the application.

## JOB MANAGEMENT & SCHEDULING

The master node keeps track of the ACR (Available CPU Resources), AMR(Available Memory Resources) and TCP/IP network information of all the nodes on the cluster using the HEPINFO program designed in shell scripting. HEPINFO program is the job manager and scheduler which compares the available computing resources of all the worker nodes and submit the job to the most available worker node on first-come first-serve basis. The users firstly login into the master node using the usernames which have been allotted to them and find the following welcome screen:

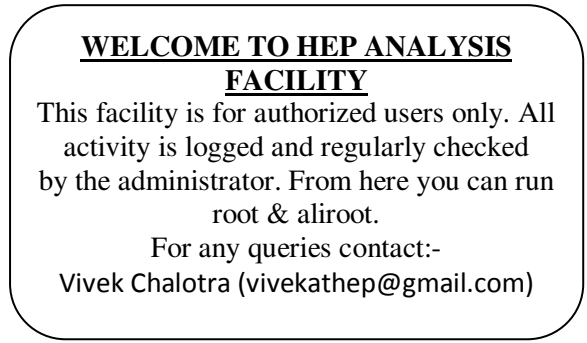

**Fig 2:** Welcome Screen

When user runs root or aliroot command here, the HEPINFO program automatically login him/her to the most available worker node. The current working directory of the user remains the same in the centralized storage i.e /Jugrid/user.

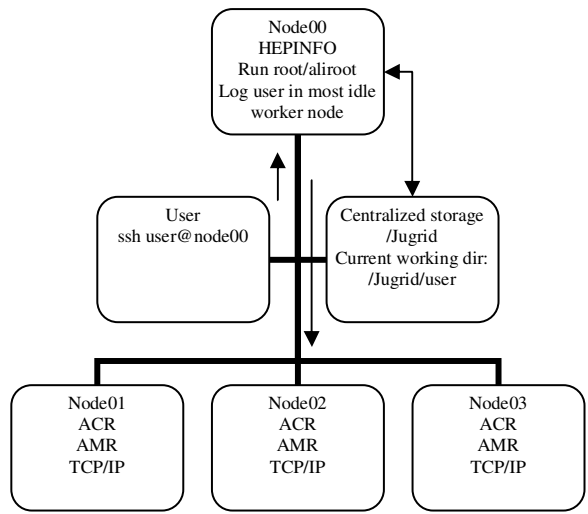

**Fig 3: Job Management**

## LIST OF CAPTIONS

| Fig 1 | : | **Overview of the Cluster** |
| Fig 2 | : | **Welcome Screen** |
| Fig 3 | : | **Job Management** |
| Table 1 | : | **Equipments Used** |